## Back to the past: on the shoulders of an academic search engine giant

**Alberto Martín-Martín[1], Enrique Orduna-Malea[2], Juan M Ayllón[1], and Emilio Delgado López-Cózar[1*]**

[1] *EC3 Research Group, Universidad de Granada, 18071 Granada, Spain*
[2] *EC3 Research Group, Polytechnic University of Valencia. Camino de Vera s/n, Valencia 46022, Spain*

* edelgado@ugr.es

**Abstract** A study released by the Google Scholar team found an apparently increasing fraction of citations to old articles from studies published in the last 24 years (1990-2013). To demonstrate this finding we conducted a complementary study using a different data source (Journal Citation Reports), metric (aggregate cited half-life), time spam (2003-2013), and set of categories (53 Social Science subject categories and 167 Science subject categories). Although the results obtained confirm and reinforce the previous findings, the possible causes of this phenomenon keep unclear. We finally hypothesize that "first page results syndrome" in conjunction with the fact that Google Scholar favours the most cited documents are suggesting the growing trend of citing old documents is partly caused by Google Scholar.

A study released by the Google Scholar team (Verstak et al 2014) finds an apparently increasing fraction of citations to old articles from studies published in the last 24 years (1990-2013). This work covers the citations from English articles published in scientific journals and conferences indexed in the 2014 release of Google Scholar Metrics. The 261 subject categories considered are grouped into 9 broad research areas. For each pair category/year and area/year group, the total number of citations and the number of citations to articles published in each preceding year are computed as well.

The authors establish three different thresholds to characterize the older articles: a) older than 10 years; b) older than 15 years; and c) older than 20 years. Thus, the percentage of citations to older documents (articles published at least 10, 15 or 20 years before the citing article) is calculated. Complementary, two periods (first half: 1990-2001; second half: 2002-2013) are set up to ascertain if the change rate in the fraction of older citations per category is either speeding up or slowing down.

The findings reveal an elevated and growing percentage of citations to old articles. In 2013, 36% of citations were to articles that are at least 10 years old; this fraction has grown 28% since 1990. The fraction of older citations increased over the period 1990-2013 in almost all scientific disciplines (231 out of 261 subject categories and 7 out of 9 broad areas); in some of them in a remarkable way (the 39% of subject categories experienced a growth over 30%). Finally, the change over the second half (19%) was significantly larger than over the first half (9%).

The study brings up with relevant questions and shows convincing results. However, the method should have provided more detailed information about the exact size of the object of study (the number of journals, articles and citations processed). Moreover, one may miss the raw data for each of the 261 subject categories, which would be helpful for the scientometrics community.

The authors employ the individual citation as the unit of observation, and refer only to journals written in English, even if they may age at different pace respect to journals written in other languages (Ruiz-Baños and Jimenez 1996). For these reasons, in such a complex phenomenon as the citation to older contributions is, providing adequate explanations about why authors are citing older material is challenging (Lariviere, Archambault, Gingras 2008), and some questions still remain unresolved.

In this sense, Verstak et al mention two broad trends that may influence the increasing fraction of older citations found. On the one hand, the dramatic growth in the number of articles published per year. On the other hand, the role that search engines are playing in finding relevant older articles, making easier for researchers to cite the most relevant articles for their work, regardless of the age of the articles. Precisely, the discussion of these two key factors shapes the main objective of this letter.

The concept of obsolescence (the phenomenon by which scientific publications are decreasing used over time) was developed within the work environment of the American College Libraries in the 1930s, concerned to keep alive library collections (separating and/or expurgating obsolete materials) in order to supply a broad education to their students (Gross and Gross 1927). It was empirically studied for the first time by Gosnell (1944). Since then, various methods have been proposed to measure the aging process of scientific literature. Among these, the most popular one is known by the name of “Half-Life”, proposed by Burton and Kebler in 1960.

In order to replicate and verify the results provided by Verstak et al, we have precisely carried out an alternative procedure to calculate the aging pace of citations by means of the aggregate cited half-life indicator from Thomson Reuters' Journal Citation Reports (JCR) for the 53 subject categories in the Social Sciences edition (Appendix I), and also for the 167 Science categories (Appendix II). Since the JCR included the aggregate cited half-life in 2003, we have used this year and compared it to the data shown in 2013. For this reason, we cannot accurately replicate the study by Verstak et al, which analyses data from 1990 onwards.

Davis and Cochran (2015) also replied Verstak et al report by analysing 4,937 journals listed in the Journal Citation Reports (1997-2013), reporting that 209 of 229 subject categories experienced increasing cited half-lifes. While they focus on the mean cited half-life and growth per annum, we analyse the differences between 2003 and 2013 providing thus a complementary view. Finally, we have not being able to analyse journals from the Arts & Humanities, since there are no cited half-life indicators for these disciplines in JCR.

As can be seen in the Appendixes I and II, these results (data from JCR) match closely with those described by Verstak et al (data from Google Scholar). Even the decrease in the cited half-life of Chemistry and Engineering subjects is obtained, already highlighted by Davis and Cochran (2015).

Despite the coincidence in the core of our analyses (the growth of citations to old literature and a decrease in Chemistry and Engineering), the reasons that explain this extension in the life cycle of scientific documents are still open to discussion.

*Exponential growth and obsolescence*

The first factor that should be considered has already been studied extensively: the relation between the exponential growth of scientific production and the pace of obsolescence. Although Price (1963) suggested this bond, it is Line (1970) who precisely described the relationship between these two phenomena, determining the following statement: if the number of published articles grows rapidly, an equally rapid growth in the number of citations to recently published articles can be expected. The faster the pace at which scientific studies are published, the faster these publications become obsolete.

However, Egghe (1993) mathematically proved how the growth of scientific production modifies the pace of obsolescence, pointing out technical differences between diachronous and synchronous studies: obsolescence increases in synchronous studies (like the work carried out by Verstak et al.) and decreases in diachronous studies (like the work carried out by Davis and Cochran). Other authors like Nakamoto (1988) claimed however the equivalence of synchronous and diachronous approaches on the ground that they generated similar curves. Anyway, the literature has systematically determined that “the growth can influence aging but that it does not cause aging” (Egghe & Rousseau 2000).

Therefore, are we entering in a period of slow scientific growth? Is today's scientific production of a lower quality, not providing as many new discoveries and techniques? Since the answers to these questions seem to be negative at first glance, we need hence further explanations – besides the scientific output growth rate - to explain the increase in citations to old documents.

*Technology and search engines*

The second factor that may explain the growth in the fraction of citations to old documents could be the recent changes in scientific communication brought about by advancements in information and communications technologies (the development of the Internet, the creation and widespread adoption of the Web). The study by Google Scholar's team heads in this direction when they call that widespread citation of older documents is now possible thanks to accessibility improvements to scientific knowledge (digitization of old documents, proliferation of repositories, and search engines), although Davis and Cochran (2015) argument against this “plain” explanation. These authors conclude that the trend to cite older papers cannot be only explained by technology (as Verstak et al assume) because this trend began decades before Google or even the Internet were invented, a thesis defended by Lariviere, Archambault and Gingras (2008) as well.

Effectively, growing cited half-life may be a result of a structural shift to fund applied research over fundamental science. As a result, fewer theoretical studies may require new authors to cite an increasingly aging literature in order to give credit to the founders. This may explain why in some categories the cited half-life decreased over the observation period (especially the general fields of Chemistry and Engineering). These results are consistent with those obtained by Verstak et al, who detect a drop in the fraction of citations to old articles for Chemical & Material Sciences and Engineering.

Notwithstanding, the technology argument seems quite reasonable, and it is supported by the changes in scientist's reading habits, previously described by Tenopir and King (2008), who found that the advent of digital technologies on searching, storing and publishing had a dramatic impact on academic information seeking and reading patterns. Scientists have substantially increased their number of readings; they read from a much broader range of sources of articles due to access to enlarged library electronic collections, on-line searching capabilities, access to other new sources such as author websites and, what is more relevant to the issue at hand: “the age of articles read appears to be fairly stable over the years, with a recent increase in reading of older articles. Electronic technologies have enhanced access to older articles.”

### *The Google Scholar effect*

Apart from the overall effect that technological advancement may have on the citations to old documents, the particular influence of Google Scholar on these changes should be discussed. Google Scholar has become the most used source for searching scientific information (Orduña-Malea et al 2014), the currently largest public source of scientific information (Orduña-Malea et al 2015; Ortega 2014), and it grows at a higher pace than its competitors (Orduña-Malea and Delgado López-Cózar 2014). So we may conclude that this search engine is contributing in a significant way. Although Verstak et al do not explicitly remark this, the use of the product's motto "Stand on the shoulders of giants" in the title of the study might be interpreted as suggesting that the growing trend of citing old documents has been in part caused by Google Scholar. This particular effect, in so far as indicative of a cognitive change in research students, has been already tested (Varshney, 2012).

And there is truth to this claim since it is undeniable that Google Scholar has revolutionized the way we search and access scientific information (Van Noorden 2014). A clear manifestation of this reflects in the way results are currently displayed, a key issue that determines how the document is accessed, read, and potentially cited. The “first results page syndrome” is probably causing that users are increasingly getting used to access only those documents that are displayed in the first results pages. In Google Scholar, as opposed to traditional bibliographic databases (Web of Science, Scopus) and library catalogues, documents are sorted by relevance and not by their publication date. And relevance - in the eyes of Google Scholar - is strongly influenced by citations (Beel and Gipp 2009; Martín-Martín et al 2014).

Google Scholar favours the most cited documents (which obviously are also the oldest documents) over more recent ones, which have had less time to accumulate citations. Although it is true that Google Scholar offers the possibility of sorting and filtering searches by publication date, this option is not used by default. Conversely, traditional databases sort their results by publication date (prioritizing novelty and recentness), allowing the user to select other criteria if they are so inclined (citation, relevance, name of first author, publication name, etc.).

Is Google Scholar contributing to change reading and citation habits because of the way information is searched and accessed through its search engine? If this is true, we could say that the work of the thousands of intellectual labourers that support with their citations the findings of the heroes of science is resting on the shoulders of just a one giant, which takes the form of an academic search engine: Google Scholar.

**Appendix I. Aggregate Cited Half-Life according to subject categories (Social Sciences. Journal Citations Reports (2003 and 2013).**

| SUBJECT CATEGORIES | 2003 | 2013 | CHANGE | TREND |
|---|---|---|---|---|
| History | >10.0 | **>10.0** | | ? |
| History & Philosophy of Science | >10.0 | **>10.0** | | ? |
| History of Social Sciences | >10.0 | **>10.0** | | ? |
| Social Sciences, Mathematical Methods | >10.0 | **>10.0** | | ? |
| Sociology | >10.0 | **>10.0** | | ? |
| Psychology, Educational | >10.0 | **>10.0** | | ? |
| Psychology, Mathematical | >10.0 | **>10.0** | | ? |
| Psychology, Social | >10.0 | **>10.0** | | ▲ |
| Economics | >10.0 | **>10.0** | | ? |
| Business | 9.9 | **>10.0** | | ▲ |
| Psychology, Multidisciplinary | 9.7 | **>10.0** | | ▲ |
| Psychology, Applied | 9.5 | **>10.0** | | ▲ |
| Business, Finance | 9.4 | **>10.0** | | ▲ |
| Psychology, Psychoanalysis | 9.3 | **>10.0** | | ▲ |
| Management | 9.1 | **>10.0** | | ▲ |
| Law | 8.5 | **>10.0** | | ▲ |
| Industrial relations & Labor | 8.4 | **>10.0** | | ▲ |
| Psychology, Biological | 9.3 | **9.7** | 0.4 | ▲ |
| Anthropology | 9.6 | **9.6** | 0.0 | — |
| Area Studies | 7.6 | **9.3** | 1.7 | ▲ |
| Demography | 8.4 | **9.2** | 0.8 | ▲ |
| Political Science | 8.3 | **9.2** | 0.9 | ▲ |
| Psychology, Developmental | 8.8 | **9.0** | 0.2 | ▲ |
| Psychology, Experimental | 9.0 | **8.8** | -.02 | ▼ |
| Psychology, Clinical | 8.2 | **8.8** | 0.6 | ▲ |
| Family Studies | 8.2 | **8.7** | 0.5 | ▲ |
| Planning & Development | 7.0 | **8.7** | 1.7 | ▲ |
| Ergonomics | 8.4 | **8.6** | 0.2 | ▲ |
| Criminology & Penology | 7.9 | **8.5** | 0.6 | ▲ |
| Communication | 8.3 | **8.4** | 0.1 | ▲ |
| Social Work | 8.0 | **8.4** | 0.4 | ▲ |
| Women's Studies | 7.2 | **8.4** | 1.2 | ▲ |
| Psychiatry | 7.7 | **8.3** | 0.6 | ▲ |
| Urban Studies | 6.5 | **8.3** | 1.8 | ▲ |
| Education & Educational Research | 8.2 | **8.2** | 0.0 | — |
| Social Sciences, Interdisciplinary | 8.5 | **8.1** | -0.4 | ▼ |
| Gerontology | 6.5 | **8.1** | 1.6 | ▲ |
| International Relations | 6.5 | **7.9** | 1.4 | ▲ |
| Public Administration | 7.2 | **7.8** | 0.6 | ▲ |
| Social Issues | 7.1 | **7.7** | 0.6 | ▲ |
| Transportation | 7.6 | **7.4** | -0.2 | ▼ |
| Geography | 6.7 | **7.4** | 0.7 | ▲ |
| Information Science & Library Science | 6.1 | **7.4** | 1.3 | ▲ |
| Ethics | 7.4 | **7.3** | -0.1 | ▼ |
| Rehabilitation | 7.1 | **7.2** | 0.1 | ▲ |
| Public, Environmental & Occupational Health | 7.0 | **7.2** | 0.2 | ▲ |
| Social Sciences, Biomedical | 6.9 | **7.2** | 0.3 | ▲ |
| Substance Abuse | 6.3 | **7.0** | 0.7 | ▲ |
| Health Policy & Services | 6.0 | **7.0** | 1.0 | ▲ |
| Education, Special | 8.8 | **6.9** | -1.9 | ▼ |
| Nursing | 6.9 | **6.9** | 0.0 | — |
| Ethnic Studies | 6.1 | **6.8** | 0.7 | ▲ |
| Environmental Studies | 7.2 | **6.5** | -0.7 | ▼ |

## Appendix II. Aggregate Cited Half-Life according to subject categories (Sciences). Journal Citations Reports (2003 and 2013).

| SUBJECT CATEGORIES | 2003 | 2013 | CHANGE | TREND |
|---|---|---|---|---|
| Mathematics | >10.0 | >10.0 | | ? |
| Statistics & Probability | >10.0 | >10.0 | | ? |
| Mineralogy | >10.0 | >10.0 | | ? |
| Omithology | >10.0 | >10.0 | | ? |
| Zoology | >10.0 | >10.0 | | ? |
| Limnology | 9.7 | >10.0 | | ▲ |
| Agriculture, Soil Science | 9.6 | >10.0 | | ▲ |
| Geochemistry & Geophysics | 9.5 | >10.0 | | ▲ |
| Paleontology | 8.9 | >10.0 | | ▲ |
| Engineering, Aerospace | 8.6 | >10.0 | | ▲ |
| Geology | 8.3 | >10.0 | | ▲ |
| Mathematics, Interdisciplinary Applications | >10.0 | 9.9 | | ▼ |
| Entomology | 9.4 | **9.7** | 0.3 | ▲ |
| Engineering, Geological | 9.3 | **9.7** | 0.4 | ▲ |
| Oceanography | 8.5 | **9.7** | 1.2 | ▲ |
| Fisheries | 8.4 | **9.6** | 1.2 | ▲ |
| Marine & Freshwater Biology | 8.7 | **9.5** | 0.8 | ▲ |
| Computer Science, Theory & Methods | 8.2 | **9.3** | 1.1 | ▲ |
| Horticulture | 7.7 | **9.3** | 1.6 | ▲ |
| Acoustics | 8.3 | **9.2** | 0.9 | ▲ |
| Physics, Atomic, Molecular & Chemical | 8.9 | **9.1** | 0.2 | ▲ |
| Mechanics | 9.8 | **9.0** | -0.8 | ▼ |
| Anatomy & Morphology | 7.6 | **9.0** | 1.4 | ▲ |
| Engineering, Ocean | 6.6 | **8.9** | 2.3 | ▲ |
| Orthopedics | 9.3 | **8.7** | -0.6 | ▼ |
| Materials Science, Ceramics | 8.2 | **8.7** | 0.5 | ▲ |
| Agronomy | 8.1 | **8.7** | 0.6 | ▲ |
| Computer Science, Hardware & Architecture | 8.1 | **8.7** | 0.6 | ▲ |
| Plant Sciences | 7.5 | **8.7** | 1.2 | ▲ |
| Physiology | 7.2 | **8.7** | 1.5 | ▲ |
| Mining & Mineral Processing | 7.1 | **8.7** | 1.6 | ▲ |
| Mathematics, Applied | 8.7 | **8.6** | -0.1 | ▼ |
| Ecology | 8.4 | **8.6** | 0.2 | ▲ |
| Agricultural Economics & Policy | 8.2 | **8.6** | 0.4 | ▲ |
| Physics, Mathematical | 6.7 | **8.6** | 1.9 | ▲ |
| Operations Research & Management Science | 9 | **8.5** | -0.5 | ▼ |
| Agriculture, Dairy & Animal Science | 8.8 | **8.5** | -0.3 | ▼ |
| Otorhinolaryngology | 8.6 | **8.5** | -0.1 | ▼ |
| Biology | 7.7 | **8.5** | 0.8 | ▲ |
| Engineering, Petroleum | >10.0 | **8.4** | | ▼ |
| Dentistry, Oral Surgery & Medicine | 8.9 | **8.4** | -0.5 | ▼ |
| Forestry | 7.7 | **8.4** | 0.7 | ▲ |
| Geosciences, Multidisciplinary | 7.7 | **8.4** | 0.7 | ▲ |
| Behavioral Sciences | 8.1 | **8.3** | 0.2 | ▲ |
| Evolutionary Biology | 7.3 | **8.3** | 1.0 | ▲ |
| Physics, Fluids & Plasmas | 6.3 | **8.3** | 2.0 | ▲ |
| Sport Sciences | 8.2 | **8.2** | 0.0 | — |
| Imaging Science & Photographic Technology | 6.7 | **8.2** | 1.5 | ▲ |
| Nuclear Science & Technology | 6.5 | **8.2** | 1.7 | ▲ |
| Biochemistry & Molecular Biology | 6.0 | **8.2** | 2.2 | ▲ |
| Microscopy | 5.6 | **8.2** | 2.6 | ▲ |
| Developmental Biology | 5.4 | **8.2** | 2.8 | ▲ |
| Engineering, Industrial | 7.5 | **8.1** | 0.6 | ▲ |
| Medicine, General & Internal | 7.0 | **8.1** | 1.1 | ▲ |
| Veterinary Sciences | 7.8 | **8.0** | 0.2 | ▲ |
| Agriculture, Multidisciplinary | 7.5 | **8.0** | 0.5 | ▲ |
| Engineering, Mechanical | 7.4 | **8.0** | 0.6 | ▲ |
| Physics, Multidisciplinar | 7.1 | **8.0** | 0.9 | ▲ |
| Anesthesiology | 6.9 | **8.0** | 1.1 | ▲ |
| Materials Science, Paper & Wood | 9.5 | **7.9** | -1.6 | ▼ |
| Computer Science, Software Engineering | 7.9 | **7.9** | 0.0 | — |
| Ophthalmology | 7..4 | **7.9** | 0.5 | ▲ |
| Remote Sensing | 7.1 | **7.9** | 0.8 | ▲ |
| Multidisciplinary Sciences | 6.9 | **7.9** | 1.0 | ▲ |
| Computer Science, Cybernetics | 6.7 | **7.9** | 1.2 | ▲ |
| Surgery | 7.6 | **7.8** | 0.2 | ▲ |
| Psychiatry | 6.9 | **7.8** | 0.9 | ▲ |
| Materials Science, Coatings & Films | 6.8 | **7.8** | 1.0 | ▲ |
| Peripheral Vascular Disease | 5.6 | **7.8** | 2.2 | ▲ |
| Medical Laboratory Technology | 7.7 | **7.7** | 0.0 | — |
| Meteorology & Atmospheric Sciences | 7.3 | **7.7** | 0.4 | ▲ |
| Pathology | 6.8 | **7.7** | 0.9 | ▲ |
| Physics, Nuclear | 6.8 | **7.7** | 0.9 | ▲ |
| Neurosciences | 6.2 | **7.7** | 1.5 | ▲ |
| Biophysics | 5.9 | **7.7** | 1.8 | ▲ |
| Water Resources | 7.6 | **7.6** | 0.0 | — |
| Dermatology | 7.2 | **7.6** | 0.4 | ▲ |
| Spectroscopy | 6.2 | **7.6** | 1.4 | ▲ |
| Geography, Physical | 7.3 | **7.5** | 0.2 | ▲ |
| Chemistry, Inorganic & Nuclear | 7.0 | **7.5** | 0.5 | ▲ |
| Computer Science, Artificial Intelligence | 6.5 | **7.5** | 1.0 | ▲ |
| Reproductive Biology | 5.8 | **7.5** | 1.7 | ▲ |
| Biodiversity Conservation | 9.2 | **7.4** | -1.8 | ▼ |
| Construction & Building Technology | 7.5 | **7.4** | -0.1 | ▼ |
| Pediatrics | 7.0 | **7.4** | 0.4 | ▲ |
| Microbiology | 5.8 | **7.4** | 1.6 | ▲ |
| Critical Care Medicine | 5.2 | **7.4** | 2.2 | ▲ |
| Crystallography | 8.3 | **7.3** | -1.0 | ▼ |
| Engineering, Multidisciplinary | 7.6 | **7.3** | -0.3 | ▼ |
| Rehabilitation | 7.2 | **7.3** | 0.1 | ▲ |
| Clinical Neurology | 6.8 | **7.3** | 0.5 | ▲ |
| Food Science & Technology | 7.4 | **7.2** | -0.2 | ▼ |
| Emergency Medicine | 6.6 | **7.2** | 0.6 | ▲ |
| Mycology | 6.3 | **7.2** | 0.9 | ▲ |
| Substance Abuse | 6.0 | **7.2** | 1.2 | ▲ |
| Respiratory System | 5.6 | **7.2** | 1.6 | ▲ |
| Cell Biology | 5.4 | **7.2** | 1.8 | ▲ |
| Metallurgy & Metallurgical Engineering | 6.9 | **7.1** | 0.2 | ▲ |
| Thermodynamics | 8.2 | **7.0** | -1.2 | ▼ |
| Transportation Science & Technology | 7.6 | **7.0** | -0.6 | ▼ |
| Robotics | 7.3 | **7.0** | -0.3 | ▼ |
| Obstetrics & Gynecology | 6.8 | **7.0** | 0.2 | ▲ |
| Polymer Science | 6.8 | **7.0** | 0.2 | ▲ |
| Chemistry, Organic | 6.7 | **7.0** | 0.3 | ▲ |
| Engineering, Electrical & Electronic | 6.7 | **7.0** | 0.3 | ▲ |
| Computer Science, Information Systems | 6.6 | **7.0** | 0.4 | ▲ |
| Medical Informatics | 6.4 | **7.0** | 0.6 | ▲ |
| Radiology Nuclear Medicine & Medical Imaging | 6.3 | **7.0** | 0.7 | ▲ |
| Astronomy & Astrophysics | 6.2 | **7.0** | 0.8 | ▲ |
| Pharmacology & Pharmacy | 6.2 | **7.0** | 0.8 | ▲ |
| Endocrinology & Metabolism | 6.0 | **7.0** | 1.0 | ▲ |
| Toxicology | 6.0 | **7.0** | 1.0 | ▲ |
| Engineering, Manufacturing | 5.9 | **7.0** | 1.1 | ▲ |
| Genetics, Heredity | 5.4 | **7.0** | 1.6 | ▲ |
| Immunology | 5.4 | **7.0** | 1.6 | ▲ |
| Engineering, Marine | >10.0 | **6.9** | | ▼ |
| Nursing | 7.0 | **6.9** | -0.1 | ▼ |
| Materials Science, Characterization & Testing | 6.8 | **6.9** | 0.1 | ▲ |
| Andrology | 6.7 | **6.9** | 0.2 | ▲ |
| Chemistry, Analytical | 6.7 | **6.9** | 0.2 | ▲ |
| Tropical Medicine | 7.8 | **6.8** | -1.0 | ▼ |
| Nutrition & Dietetics | 6.4 | **6.8** | 0.4 | ▲ |
| Chemistry, Applied | 6.2 | **6.8** | 0.6 | ▲ |
| Materials Science, Composites | 6.2 | **6.8** | 0.6 | ▲ |
| Computer Science, Interdisciplinary Applica | 6.0 | **6.8** | 0.8 | ▲ |
| Health Care Sciences & Services | 6.0 | **6.8** | 0.8 | ▲ |
| Medicine, Research & Experimental | 6.0 | **6.8** | 0.8 | ▲ |
| Cardiac & Cardiovascular Systems | 5.8 | **6.8** | 1.0 | ▲ |
| Automation & Control Systems | 7.8 | **6.7** | -1.1 | ▼ |
| Physics, Condensed Matter | 6.7 | **6.7** | 0.0 | — |
| Environmental Sciences | 6.5 | **6.7** | 0.2 | ▲ |
| Urology & Nephrology | 5.6 | **6.7** | 1.1 | ▲ |
| Engineering, Civil | 8.0 | **6.6** | -1.4 | ▼ |
| Geriatrics & Gerontology | 5.8 | **6.6** | 0.8 | ▲ |
| Engineering, Chemical | 7.4 | **6.5** | -0.9 | ▼ |
| Materials Science, Textiles | 7.4 | **6.5** | -0.9 | ▼ |
| Virology | 5.7 | **6.5** | 0.8 | ▲ |
| Hematology | 5.4 | **6.5** | 1.1 | ▲ |
| Biotechnology & Applied Microbiology | 5.3 | **6.5** | 1.2 | ▲ |
| Instruments & Instrumentation | 6.1 | **6.4** | 0.3 | ▲ |
| Chemistry, Medicinal | 6.0 | **6.4** | 0.4 | ▲ |
| Rheumatology | 5.8 | **6.4** | 0.6 | ▲ |
| Gastroenterology & Hepatology | 5.7 | **6.4** | 0.7 | ▲ |
| Neuroimaging | 4.9 | **6.4** | 1.5 | ▲ |
| Biochemical Research Methods | 6.8 | **6.3** | -0.5 | ▼ |
| Engineering, Biomedical | 6.2 | **6.3** | 0.1 | ▲ |
| Allergy | 5.7 | **6.3** | 0.6 | ▲ |
| Medicine, Legal | 5.3 | **6.3** | 1.0 | ▲ |
| Optics | 6.6 | **6.2** | -0.4 | ▼ |
| Oncology | 5.4 | **6.2** | 0.8 | ▲ |
| Transplantation | 5.3 | **6.2** | 0.9 | ▲ |
| Infectious Diseases | 5.2 | **6.2** | 1.0 | ▲ |
| Telecommunications | 6.5 | **6.1** | -0.4 | ▼ |
| Physics, Particles & Fields | 4.8 | **6.1** | 1.3 | ▲ |
| Electrochemistry | 6.7 | **6.0** | -0.7 | ▼ |
| Physics, Applied | 6.0 | **6.0** | 0.0 | — |
| Engineering, Environmental | 6.2 | **5.9** | -0.3 | ▼ |
| Integrative & Complementary Medicine | 5.5 | **5.9** | 0.4 | ▲ |
| Chemistry, Physical | 5.4 | **5.7** | 0.3 | ▲ |
| Chemistry, Multidisciplinary | 7.2 | **5.6** | -1.6 | ▼ |
| Medical Ethics | 5.7 | **5.5** | -0.2 | ▼ |
| Materials Science, Multidisciplinary | 5.6 | **5.4** | -0.2 | ▼ |
| Parasitology | 6.9 | **5.2** | -1.7 | ▼ |
| Materials Science, Biomaterials | 5.6 | **5.1** | -0.5 | ▼ |
| Agricultural Engineering | 7.7 | **5.0** | -2.7 | ▼ |
| Energy & Fuels | 7.0 | **4.7** | -2.3 | ▼ |